\documentclass{ws-procs9x6}
\usepackage{graphicx}

\newcommand{\nuc}[2]{${}^{#2} \rm #1$}
\newcommand{\aap}{{\it Astron. Astrophys.} }

\newcommand{\apj}{{\it Astrophys. J.} }

\newcommand{\nat}{{\it Nature} }

\newcommand{\apjs}{{\it Astrophys.~J.~S.} }

\newcommand{\ptp}{{\it Prog.~Theor.~Phys.} }
\newcommand{\pasj}{{\it Publ.~Astron.~Soc.~Japan} }

\usepackage[usenames]{color}

\def\AKA#1#2{%
  \textcolor{blue}{#1}
  \textcolor{red}{\setbox0\hbox{#2}\leavevmode
  \raise1mm\hbox to0pt{\vrule width\wd0 depth0pt height0.2mm
  \hss}\box0}}
\begin{document}

\title{
Nucleosynthesis inside Gamma-Ray Burst Accretion Disks
}

\author{Shin-ichirou Fujimoto}

\address{Department of Electronic Control, 
Kumamoto National College of Technology, Kumamoto 861-1102, Japan;
E-mail: fujimoto@ec.knct.ac.jp
}
 
\author{Masa-aki Hashimoto}

\address{
Department of Physics, School of Sciences, 
Kyushu University, Fukuoka 810-8560, Japan}

\author{Kenzo Arai and Ryuichi Matsuba}

\address{Department of Physics, Kumamoto University, Kumamoto 860-8555, Japan}

\maketitle

\abstracts{
We investigate nucleosynthesis inside both a
gamma-ray burst accretion disk and 
a wind launched from an inner region of the disk 
using one-dimensional models of the disk and wind 
and a nuclear reaction network.
Far from a central black hole, 
the composition of accreting gas is taken to be that of 
an O-rich layer of a massive star before core collapse.
We find that 
the disk consists of 
five layers characterized by dominant elements:
\nuc{O}{16}, \nuc{Si}{28}, \nuc{Fe}{54} (and \nuc{Ni}{56}), 
\nuc{He}{4}, and nucleons, 
and the individual layers shift inward with keeping the overall
profiles of compositions as the accretion rate decreases. 
\nuc{Ni}{56} are abundantly ejected through the wind
from the inner region of the disk
with the electron fraction $\simeq 0.5$.
In addition to iron group, 
elements heavier than Cu, in particular \nuc{Cu}{63} and \nuc{Zn}{64}, 
are massively produced through the wind.
Various neutron-rich nuclei can be also produced in 
the wind 
from neutron-rich regions of the disk, 
though the estimated yields have large uncertainties.
}

\section{Introduction}

Observational evidences have been accumulated 
for a connection between gamma-ray bursts (GRBs) and
supernovae (SNe):
association of SN 1998bw and GRB 980425\cite{Ga98}
and SN 2003dh
in afterglow of GRB 030329.\cite{Hj03}
A {\itshape collapsar} model is one of promising scenarios 
to explain a huge gamma-ray production in GRBs and
GRB/SN connections.\cite{MW99,MWH02}
During collapse of massive stars, 
stellar material greater than several solar masses 
falls back on a new-born black hole
with extremely high accretion rates ($\le 1  M_\odot\,$s$^{-1}$).\cite{WW95}
An accretion disk forms around the hole due to the angular momentum
of the fallback material.\cite{MW99,MWH02,MNHNS97}
In the context of the collapsar model, 
jet-like explosion driven by neutrino annihilation and 
nucleosynthesis in the jet has been investigated.\cite{NMYTS03}
Although \nuc{Ni}{56} with high velocity ($> 0.1 c$) can be massively produced 
, it is not sufficient for an observed amount in SN 1998bw.\cite{NMYTS03}
In addition to the  production  via the jet, 
massive synthesis of \nuc{Ni}{56} is also suggested 
in winds launched from the accretion disk.\cite{MW99,PWH03}
Neutron-rich nuclei may be produced through r-process
inside the wind ejected from an inner, neutron-rich region of the disk.\cite{PWH03}
In the present paper, 
we examine nucleosynthesis inside a GRB accretion disk and
investigate abundance change through the wind launched from the disk.

\section{Disk Model and Input Physics}

We construct a steady, axisymmetric model\cite{DPN02,KM02} of the disk 
around a black hole of mass $M$ with accretion rates 
up to $10 M_\odot\,$s$^{-1}$.
The black hole mass is fixed to be 3 $M_\odot$ and the viscosity 
parameter is set to be $\alpha_{\rm vis} = 0.1$.
Figure 1 shows the profiles of density, $\rho$, (thick lines) 
and temperature, $T$, (thin lines)  in the accretion disk 
for 
$\dot{M} = 1$ (solid lines), 
$0.1$ (dotted lines), and
$0.01 M_{\odot} \rm \,s^{-1}$ (dashed lines).
These profiles are roughly agreement with the corresponding profiles
in another disk model\cite{DPN02}
except for the region near the disk inner edge; 
our density and temperature drop rapidly near the inner edge.
This is attributed to our use of 
the pseudo-Newtonian potential and the zero torque condition 
at the  inner edge.

\begin{figure}[htbp]
\includegraphics[angle=-90,width=10.7cm]{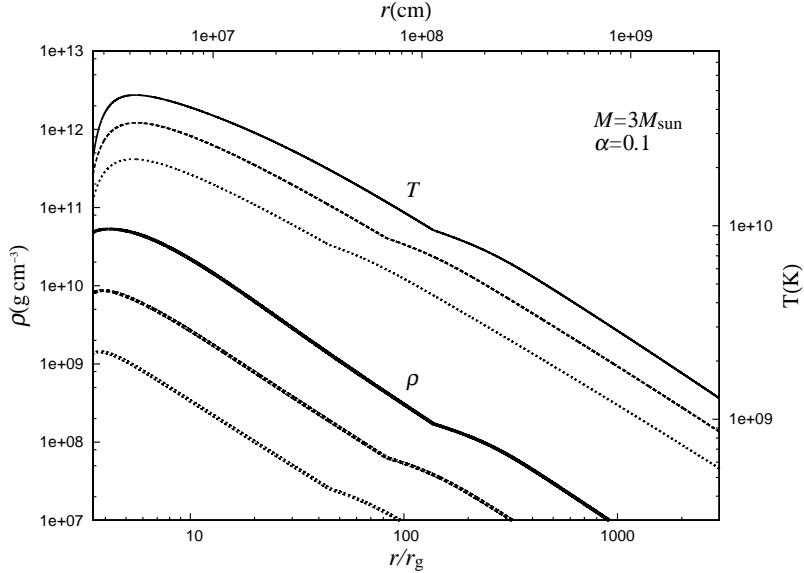}
\caption{Density and temperature profiles inside the disk with 
0.01, 0.1 and $1 M_\odot\,$s$^{-1}$.
}
\end{figure}

Chemical composition of accreting gas far from the black hole
is set to be that of an O-rich layer of a 20 $M_{\odot}$ 
star before core collapse.\cite{H95}
Once temperatures and densities are evaluated inside the disk,
using a nuclear reaction network,\cite{F01,F03}
which includes 463 nuclei up to $\rm {}^{94}Kr$, 
we follow evolution of the composition in the accreting material
during the infall onto the hole.
We note that at the inner region of the disk, 
where $T \ge 9 \times 10^9 \rm K$, 
the composition is in nuclear statistical equilibrium (NSE).\cite{C68}

\section{Abundance Distribution inside Disks}

Figure 2 shows the abundance profiles of representative nuclei 
inside the disk with  $\dot{M}=0.1 M_{\odot}\,$s$^{-1}$.
Far from the black hole $r > 1000 r_{\rm g}$, 
where $r_{\rm g} = 2GM/c^2$ is the Schwarzschild radius, 
accreting gas keeps presupernova composition, or that of the O-rich layer, 
because of low temperatures ($T < 2 \times 10^9 \rm K$). 
As the material falls down, 
the gas becomes rich in iron-group elements
via explosive O-burning, followed by Si-burning. 
The processed heavy elements, however, are destroyed to helium, 
and finally to protons and neutrons through photodisintegrations
deep inside the disk. 
Near the inner edge of the disk, 
neutrons are the most abundant by electron captures on protons.
Considerable amounts of D, T and $^6$Li exist due to NSE.

\begin{figure}[htbp]
\includegraphics[angle=-90,width=10.7cm]{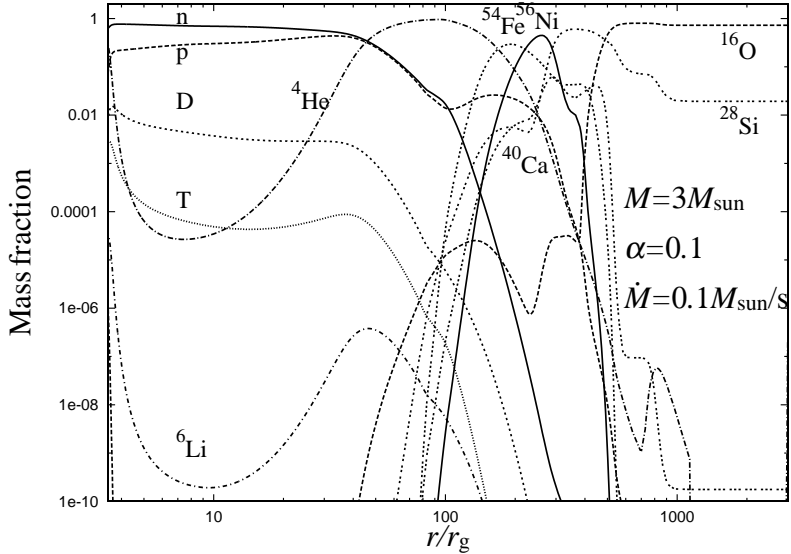}
\caption{
Abundance profiles of representative nuclei inside the disk
with $0.1 M_{\odot} \rm \,s^{-1}$. 
}
\end{figure}

For higher accretion rates, the density is higher,
the electron capture is more efficient, 
and consequently the disk becomes more neutron-rich. 
In fact, the ratios of neutron to proton are 1.33, 3.50 and 10.4 
for $\dot{M}$ = 0.01, 0.1 and 1 $ M_\odot\,$s$^{-1}$, 
respectively, near the inner edge of the disk.
Radial profiles of the electron fraction are similar to 
those in the other authors.\cite{PWH03}

\begin{figure}[htbp]
\includegraphics[angle=-90,width=10.7cm]{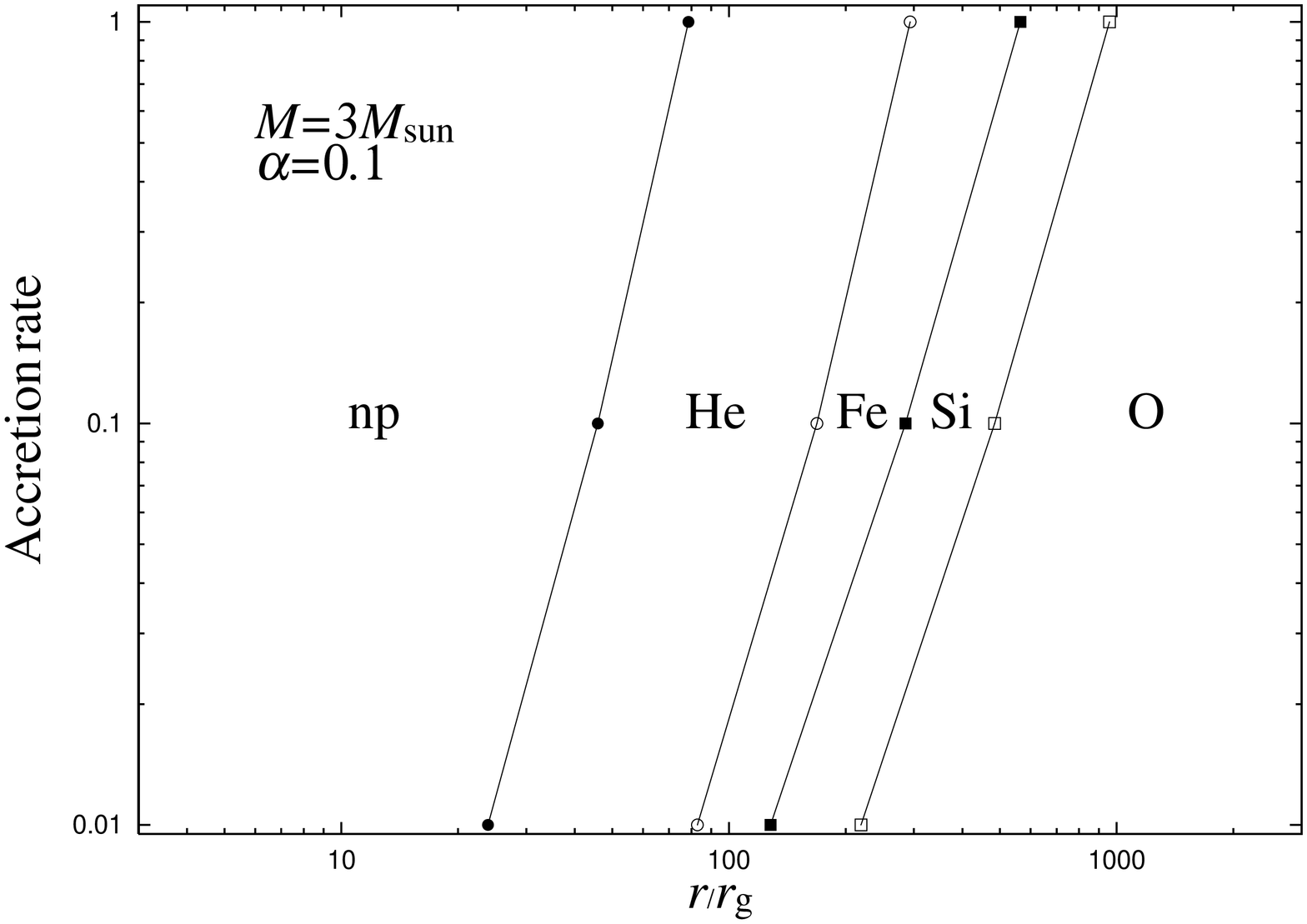}
\caption{
Interfaces of various  layers .
The disk consists of five layers characterized by dominant elements:
\nuc{O}{16}, \nuc{Si}{28}, \nuc{Fe}{54} (and \nuc{Ni}{56}), 
\nuc{He}{4}, and nucleons.
}
\end{figure}

The disk consists of five layers characterized by dominant elements:
\nuc{O}{16}, \nuc{Si}{28}, \nuc{Fe}{54} (and \nuc{Ni}{56}), \nuc{He}{4}, and nucleons.
Hereafter, these five layers are referred to as 
the O-rich, Si-rich, Fe-rich, He-rich, and np-rich disk layers.
Figure 3 shows the interfaces of these layers
for $\dot{M} = 0.01 - 1 M_\odot\,$s$^{-1}$.
Temperatures increase with the increasing accretion rate 
at a given radius,
so that the individual layers shift outward, 
but the overall profiles of composition are preserved.
This is because nucleosynthesis inside a disk mainly depends 
on the temperature distribution of the disk.
Averaged mass fractions over the individual layers 
of 40 abundant nuclei are given in Table 1
for the Fe-rich and He-rich disk layers with
$\dot{M} = $ 0.01 and 0.1 $ M_\odot\,$s$^{-1}$.
We find that the averaged abundances of the individual disk layers 
are not significantly changed as the accretion rates.

\begin{table}[htbp]
\tbl{Averaged abundances in the Fe and He layer of 
the disk before decay.}
{
\begin{tabular} {|cccc|cccc|} \hline
%%%%%
 \multicolumn{4}{|c|}{Fe-rich disk layer} &
 \multicolumn{4}{c|} {He-rich disk layer} \cr \hline
 \multicolumn{2}{|c}{$0.01 M_{\odot}\rm \,s^{-1}$} &
 \multicolumn{2}{c|}{$0.1 M_{\odot}\rm \,s^{-1}$} &
 \multicolumn{2}{c}{$0.01 M_{\odot}\rm \,s^{-1}$} &
 \multicolumn{2}{c|}{$0.1 M_{\odot}\rm \,s^{-1}$} \cr
 elem. & X & elem. & X & elem. & X & elem. & X \cr \hline
%%%%%
     Fe54 & 2.47E-01 & Ni56 & 2.59E-01 &  He4 & 6.68E-01 &  He4 & 6.29E-01 \cr
     Ni56 & 1.67E-01 & Fe54 & 1.82E-01 & Fe54 & 5.12E-02 & Fe54 & 5.49E-02 \cr
     Co55 & 1.13E-01 & Co55 & 1.17E-01 &    p & 4.34E-02 &    p & 4.08E-02 \cr
     Ni58 & 7.22E-02 & Ni58 & 6.10E-02 & Fe56 & 3.58E-02 & Fe56 & 3.60E-02 \cr
      He4 & 6.77E-02 & Ni57 & 5.12E-02 &    n & 3.18E-02 &    n & 3.04E-02 \cr

     Ni57 & 4.58E-02 & Si28 & 4.68E-02 & Fe55 & 2.68E-02 & Fe55 & 3.02E-02 \cr
     Si28 & 4.28E-02 &  He4 & 4.58E-02 & Mn53 & 1.88E-02 & Mn53 & 2.13E-02 \cr
      S32 & 3.88E-02 &  S32 & 4.18E-02 & Co57 & 1.40E-02 & Co57 & 1.62E-02 \cr
     Fe52 & 2.41E-02 & Fe52 & 3.26E-02 & Cr52 & 1.37E-02 & Cr52 & 1.49E-02 \cr
     Fe53 & 2.19E-02 & Fe53 & 2.22E-02 & Ni58 & 1.27E-02 & Ni58 & 1.45E-02 \cr

     Ca40 & 1.95E-02 & Ca40 & 2.12E-02 & Mn54 & 1.07E-02 & Mn54 & 1.41E-02 \cr
     Fe55 & 1.79E-02 & Ar36 & 1.82E-02 & Cr50 & 6.81E-03 & Cr50 & 8.35E-03 \cr
        p & 1.75E-02 & Co56 & 1.47E-02 & Co56 & 5.92E-03 & Co56 & 7.81E-03 \cr
     Ar36 & 1.72E-02 &    p & 1.33E-02 & Co55 & 5.62E-03 & Co55 & 7.31E-03 \cr
     Co56 & 1.64E-02 & Fe55 & 1.31E-02 & Ni59 & 5.56E-03 & Ni59 & 6.94E-03 \cr

     Cr50 & 1.38E-02 & Cr50 & 1.03E-02 & Ni60 & 4.92E-03 & Cr51 & 5.57E-03 \cr
     Co57 & 1.08E-02 & Mn51 & 9.36E-03 & Cr51 & 4.27E-03 & Ni60 & 5.50E-03 \cr
     Mn51 & 9.54E-03 & Co57 & 8.30E-03 & Fe57 & 4.18E-03 & Mn55 & 5.28E-03 \cr
     Mn53 & 9.46E-03 & Mn53 & 6.85E-03 & Mn55 & 4.16E-03 & Fe57 & 5.22E-03 \cr
     Fe56 & 5.96E-03 & Fe56 & 3.84E-03 & Co58 & 2.99E-03 & Co58 & 4.05E-03 \cr

     Ni59 & 4.48E-03 & Ni59 & 3.74E-03 & Mn52 & 2.62E-03 & Mn52 & 3.85E-03 \cr
     Mn52 & 3.65E-03 & Mn52 & 3.29E-03 & Ni57 & 2.33E-03 & Ni57 & 3.20E-03 \cr
     Cr48 & 2.49E-03 & Cr48 & 3.07E-03 & Cr53 & 2.22E-03 & Fe53 & 2.92E-03 \cr
     Cr49 & 2.05E-03 & Cr49 & 1.99E-03 & Fe53 & 1.98E-03 & Cr53 & 2.89E-03 \cr
     Cr51 & 1.20E-03 & Cr51 & 9.56E-04 &  V49 & 1.35E-03 &  V49 & 2.04E-03 \cr

     Ni60 & 1.12E-03 &  V47 & 8.37E-04 &  V51 & 1.32E-03 & Mn51 & 1.88E-03 \cr
     Cr52 & 1.03E-03 & Ni60 & 8.04E-04 & Co59 & 1.32E-03 &  V51 & 1.74E-03 \cr
      V47 & 8.03E-04 & Cu59 & 7.94E-04 & Fe58 & 1.29E-03 & Co59 & 1.73E-03 \cr
     Cu59 & 6.08E-04 & Cr52 & 6.81E-04 & Mn51 & 1.21E-03 & Fe58 & 1.70E-03 \cr
     Mn54 & 6.08E-04 & Co54 & 5.66E-04 & Si28 & 1.04E-03 & Si28 & 1.38E-03 \cr

      K39 & 3.85E-04 & Mn54 & 5.03E-04 & Ni56 & 9.70E-04 & Ni56 & 1.35E-03 \cr
     Co54 & 3.81E-04 & Ti44 & 3.80E-04 & Cr54 & 9.19E-04 &  V50 & 1.28E-03 \cr
     Ti46 & 3.51E-04 & Cu60 & 3.71E-04 &  V50 & 7.93E-04 & Cr54 & 1.22E-03 \cr
     Ti44 & 3.45E-04 &  K39 & 2.92E-04 &  S32 & 7.85E-04 &  S32 & 1.04E-03 \cr
     Cu60 & 3.11E-04 & Ti46 & 2.81E-04 & Ni61 & 7.63E-04 & Ni61 & 1.03E-03 \cr

     Co58 & 3.03E-04 & Co58 & 2.61E-04 & Ti48 & 6.17E-04 &  V48 & 9.49E-04 \cr
      V48 & 2.43E-04 &  V48 & 2.59E-04 & Mn56 & 5.64E-04 & Mn56 & 9.44E-04 \cr
     Cl35 & 2.14E-04 &  V49 & 1.75E-04 &  V48 & 5.11E-04 & Ti48 & 8.85E-04 \cr
      V49 & 1.95E-04 & Ti45 & 1.59E-04 & Ti46 & 4.36E-04 & Cr49 & 6.41E-04 \cr
     Cu61 & 1.51E-04 & Cl35 & 1.56E-04 & Cr49 & 3.88E-04 & Ti46 & 6.29E-04 \cr

\hline
 \end{tabular}
}
\end{table}

\section{Chemical Composition of Disk Winds}

\subsection{\nuc{Ni}{56} in Winds}

\begin{figure}[htbp]
\includegraphics[angle=-90,width=10.7cm]{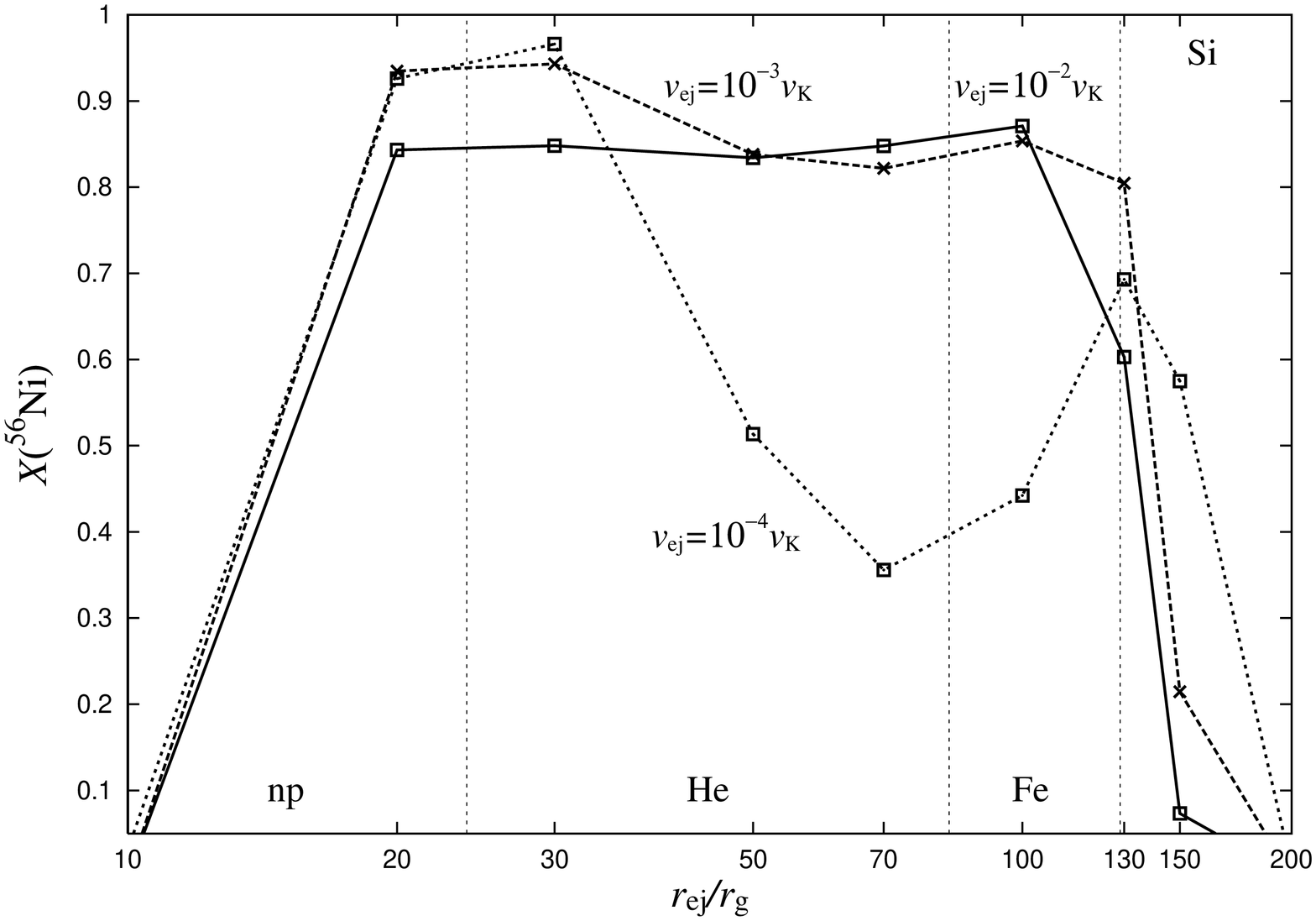}
\caption{
Mass fraction of \nuc{Ni}{56} of winds
ejected from various disk radii
and various ejection velocities.
}
\end{figure}

During accretion onto a black hole, 
some fractions of accreting gas could be ejected 
through winds from an accretion disk.
Possible processes driving the wind are 
magnetical centrifugal force\cite{DM02} and viscosity.\cite{MW99}
Abundances of ejecta through the wind change via
decay processes as well as charged particle and capture processes
because of high densities and temperatures of the ejecta.
The abundances hence depend not only on initial conditions of the ejecta
(abundances, density, temperature, and so on) 
but also on hydrodynamics of the wind.
Detailed dynamics of the wind is, however, still uncertain.
We therefore adopt a simple hydrodynamical model of the wind,\cite{FRRKTKP99}
where the gasses are assumed to be adiabatic and freely expanding.
The ejection velocity $v_{\rm ej}$ 
is set to be $10^{-4}-0.1 v_{\rm K}$ with 
the Keplerian velocity $v_{\rm K}$ 
at the ejection radius $r_{\rm ej}$ of the disk.
The density and temperature of the wind are initially 
taken to be those of the disk at $r_{\rm ej}$.
We note that the entropy per baryon are 10--20 
in units of the Boltzmann constant and that
the non-relativistic entropies are always larger than 
the relativistic ones in the ejecta.
The adiabatic index is accordingly taken to be 5/3.

Using density and temperature evolution  
calculated with the above wind model, 
we can evaluate change in abundances of the ejecta 
from an initial composition 
which is the same as in the accretion disk at $r_{\rm ej}$, 
with similar post-processing calculations to those in accretion disks.
We have calculated abundance change through wind launched from 
the inner region ($r_{\rm ej} \le 200 r_g$) of the disk with 
 $\dot{M} = 10^{-4} - 0.1  M_\odot\,$s$^{-1}$.
In briefly, the abundances through the winds ejected 
from the O-rich and Si-rich disk layers are found to be 
not largely changed from those of the disk. 
On the other hand, for the winds from the He-rich and np-rich disk layers,  
the composition is largely altered from that of the disk.

In Figure 4, we show the mass fractions of \nuc{Ni}{56} 
inside the winds from the accretion disk with $\dot{M} = 0.01 M_{\odot}\rm \,s^{-1}$.
The abscissa is the radius from which the wind is launched.
The abundances are evaluated at the time when
$T_{\rm ej} \simeq 5 \times 10^8 \rm K$. 
Abundant radioactives have not decayed significantly until the epoch.
We find that the winds from the np-rich disk layer
are abundant in \nuc{Ni}{56}, as suggested by the several authors.\cite{MW99,PWH03}
We also find that \nuc{Ni}{56} is the most abundant in the ejecta
from not only the np-rich layer but also 
the He-rich and Fe-rich layers and inner parts of the Si -rich layer.
The smallness of \nuc{Ni}{56} in the ejecta from $10 r_{\rm g}$
is attributed to efficient neutron capture on iron peak elements.
The electron fraction, $Y_{\rm e}$, is initially 0.4830 in the ejecta.
\nuc{Ni}{56} is consequently largely depleted through neutron capture, 
while more neutron rich nuclei, such as \nuc{Ni}{58}, \nuc{Ni}{60}, 
and \nuc{Zn}{64}, are abundantly synthesized.
On the other hand, in the ejecta launched from $20 r_{\rm g}$
($Y_{\rm e}$ is 0.4994 initially), 
neutron capture on iron peak elements is less efficient in the ejecta.
The mass fraction of neutrons in the ejecta from $10 r_{\rm g}$
is larger than that of protons by 0.034, 
which is too small to conduct r-process successfully.

\subsection{Chemical Composition of Winds}

\begin{figure}[htbp]
\includegraphics[angle=-90,width=10.7cm]{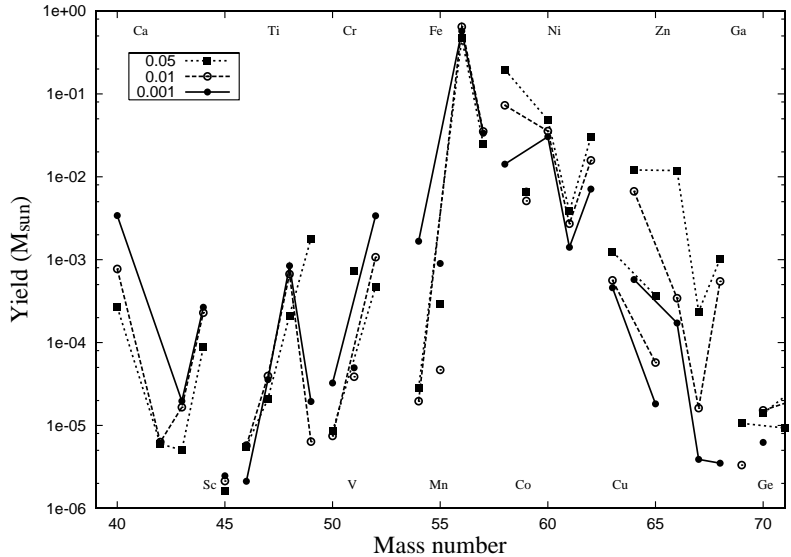}
\caption{
Yields through the wind ejected from 
the disk with 
0.001 (filled circles with solid line), 0.01 (circles with dashed line), and 
$0.05 M_{\odot}\rm \,s^{-1}$ (squares with dotted line) 
after decay
when the total mass through the wind is $1 M_{\odot}$.
}
\end{figure}

We average abundances in winds ejected from the inner region ($r \le 50 r_g$) 
of the disk to estimate the yields through the winds from the disk.
The averaging procedure is the same as in our previous study\cite{F03}
but included with a weight of $1 - \exp(-r_{\rm ej}/50r_g)$,
(or 0, whichever is smaller than 0 ), which means more massive ejection 
from smaller disk radius less than $50 r_g$.
Figure 5 shows the yields via the winds from the disk with 
$\dot{M} =$ 0.001, 0.01, and 0.05$M_{\odot}\rm \,s^{-1}$.
The total mass of ejecta is set to be $1 M_{\odot}$.\cite{MW99}
The profiles of yields are similar for different $\dot{M}$, but 
the ejected masses heavier than Cu are larger for higher $\dot{M}$.
This is due to the neutron richness of inner regions of the disk.
It should be noted that appreciable 
amounts of \nuc{Cu}{63} and \nuc{Zn}{64} can be produced through the winds
from the disk without neutron-rich regions (for 0.001$M_{\odot}\rm \,s^{-1}$).

\subsection{Chemical Composition of Winds from Neutron-rich Regions of Disks}

\begin{figure}[htbp]
\includegraphics[angle=-90,width=10.7cm]{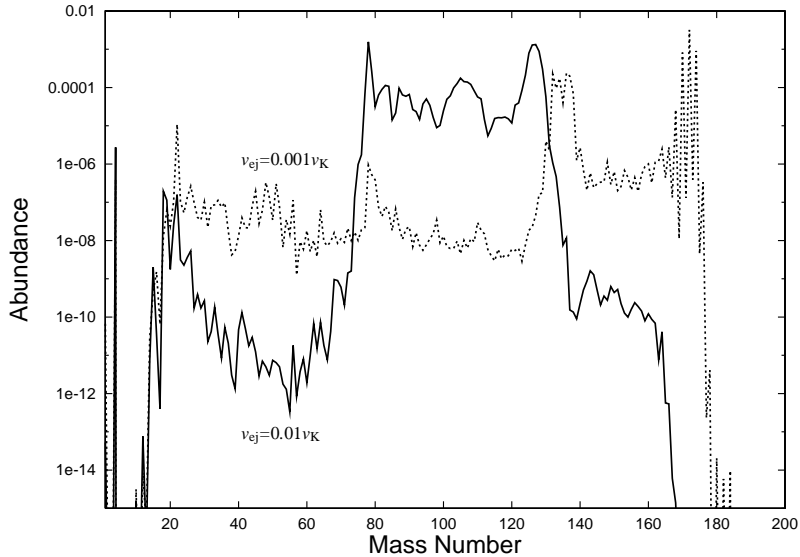}
\caption{
Abundances in the wind ejected from $r_{\rm ej} = 10r_{g}$ of the disk
with $ \dot{M} = 0.1 M_{\odot}\rm \,s^{-1}$.
The ejection velocities of the wind are taken to be $0.01 v_{\rm K}$ 
(solid line) and $0.001 v_{\rm K}$ (dotted line). 
}
\end{figure}

We investigate abundances in winds 
from neutron-rich regions of the disk with $\dot{M} \ge 0.1 M_{\odot}\rm \,s^{-1}$.
We use the same procedure as in \S 4.1
but with the larger nuclear reaction network which includes 
various neutron-rich nuclei. 
Figure 6 shows the abundances in the wind after decays 
for the representative case: 
the wind launched from $r_{\rm ej} = 10 r_{\rm g}$ of the accretion disk 
with  $\dot{M} = 0.1 M_{\odot}\rm \,s^{-1}$.
The electron fraction and the entropy per baryon are 
initially 0.30 and 13.1$k_{\rm B}$, respectively.
The ejection velocities of the wind are taken to be $0.01 v_{\rm K}$ 
(solid line) and $0.001 v_{\rm K}$ (dotted line). 
We find that neutron-rich nuclei are abundantly synthesized in the winds.
More massive nuclei can be produced in the slower wind.
This is because slower winds have enough time for nuclei 
to capture neutrons.
Nuclei with mass numbers $\simeq 120$ and 180 are abundant 
for cases with $v_{\rm ej} = 0.01 v_{\rm K}$ and $0.001 v_{\rm K}$, 
respectively.
For faster winds than $0.1 v_{\rm K}$, on the other hand, light elements are dominant.
The abundances in the winds from neutron-rich regions of the disk 
are sensitive to the ejection velocity as well as the electron fraction.
Although the electron fractions are comparable to or less than that of 
the disk in our wind model, 
the fractions may increase to be larger than 0.5 if the wind is driven 
via viscosity.\cite{PTH04}
The abundances are likely to strongly depend on hydrodynamics of the wind.
The yields from the neutron-rich disk therefore are still uncertain.

\section{Summary}

We have investigated nucleosynthesis inside the accretion disk
associated with GRBs and inside winds launched from 
an inner region of the disk using the one-dimensional disk 
and wind models and the nuclear reaction network.
The initial composition of accreting gas is taken to be that of 
an O-rich layer of a 20 $M_{\odot}$ star before the core collapse.
We have found that 
the disk consists of five layers characterized by dominant elements:
\nuc{O}{16}, \nuc{Si}{28}, \nuc{Fe}{54} (and \nuc{Ni}{56}), 
\nuc{He}{4}, and nucleons, 
and the individual layers shift inward with keeping the overall
profiles of compositions as the accretion rate decreases. 
\nuc{Ni}{56} are abundantly ejected through the wind
from the Fe-rich, He-rich and nucleon-rich disk layers
with the electron fraction $\simeq 0.5$.
In addition to iron group elements, 
heavier elements than Cu, in particular \nuc{Cu}{63} and \nuc{Zn}{64}, 
are massively produced via the wind.
Various neutron-rich nuclei can be produced through the wind 
from neutron-rich regions of the disk in our simple wind model, 
though the estimated yields have large uncertainties.

\end{document}